\documentclass[12pt]{iopart}

\usepackage{cite}

\usepackage{graphicx}

\usepackage{dcolumn}

\begin{document}

\title[Particle in a cubic box]{On the symmetry of the quantum-mechanical particle in a cubic box}

\author{Francisco M. Fern\'andez}

\address{INIFTA (UNLP, CCT La Plata-CONICET), Blvd. 113 y 64 S/N, \\
Sucursal 4, Casilla de Correo 16, 1900 La Plata,
Argentina}\ead{fernande@quimica.unlp.edu.ar}

\maketitle

\begin{abstract}
In this paper we show that the point-group (geometrical) symmetry is
insufficient to account for the degeneracy of the energy levels of the
particle in a cubic box. The discrepancy is due to hidden (dynamical
symmetry). We obtain the operators that commute with the Hamiltonian one and
connect eigenfunctions of different symmetries. We also show that the
addition of a suitable potential inside the box breaks the dynamical
symmetry but preserves the geometrical one.The resulting degeneracy is that
predicted by point-group symmetry.
\end{abstract}

\section{Introduction}

\label{sec:intro}

The particle in a one-dimensional box with impenetrable walls is
one of the first models discussed in most introductory books on
quantum mechanics and quantum chemistry\cite{CDL77,P68}. It is
suitable for showing how energy quantization appears as a result
of certain boundary conditions. Once we have the eigenvalues and
eigenfunctions for this model one can proceed to two-dimensional
boxes and discuss the conditions that render the Schr\"{o}dinger
equation separable\cite{CDL77}. The particular case of a square
box is suitable for discussing the concept of
degeneracy\cite{CDL77}. The next step is the discussion of a
particle in a three-dimensional box and in particular the cubic
box as a representative of a quantum-mechanical model with high
symmetry\cite{P68}. This model is also suitable for discussing the
perfect gas in statistical mechanics\cite{MQ76}. The spherical box
is also of great pedagogical value because it enables us to
discuss the conservation of the angular momentum of the
particle\cite{TH75}.

In two enlightening articles Leyvraz et al\cite{LFLA97} and lemus
et al\cite {LFAL98} discussed the accidental degeneracy and hidden
symmetry of the particle in square and rectangular wells,
respectively. They showed that point group (geometrical) symmetry
is insufficient to account for the degree of degeneracy of those
quantum-mechanical models and then proceeded to find the operator
that commutes with the Hamiltonian one and connects eigenfunctions
of different symmetry. In this way they constructed a larger
symmetry group that accounts for the full degeneracy of the square
well. Such an analysis had been carried out earlier by
Shaw\cite{S74}.

Leyvraz et al\cite{LFLA97} briefly commented on the problem of a
particle in a cubic box with impenetrable walls. They mentioned
that the suitable symmetry point group is $O_{h}$ and that in this
case there should be two additional dynamical symmetries. The
purpose of this paper is to pursue the study of this
quantum-mechanical model much in the same way those authors did in
the case of the two-dimensional wells. After we finished the first
draft of the present paper\cite{F13} a detailed analysis of the
symmetry of the particle in a cubic box appeared\cite{HCL13}.
However, since the focus of our paper is mainly pedagogical and
contains some results not discussed by Hern\'andez-Castillo and
Lemus\cite{HCL13} we deem that it may still be of interest.

In Sec.~\ref{sec:cubic box} we summarize the well known results of
the Schr\"{o}dinger equation for the cubic box and discuss the
degeneracy of the energy levels. In Sec.~\ref{sec:point-group
symmetry} we analyze the problem from the point of view of its
point-group symmetry and show the accidental degeneracies that
cannot be explained by the geometrical symmetry of the cube. We
extend the analysis of Leyvraz et al\cite{LFLA97} to the three
dimensional case and derive three operators that commute with the
Hamiltonian of the system and connect functions belonging to
different irreducible representations. To this end, we make
extensive use of the group projection operators\cite{T64,C90}. In
Sec.~\ref{sec:perturbation theory} we add a perturbation potential
that is invariant under the symmetry operations of the $O_{h}$
point group and show that it breaks the dynamical symmetry of the
system while conserving the geometrical one. We also discuss
another perturbation that conserves both types of symmetry.
Finally, in Sec.~\ref {sec:conclusions} we summarize the main
results and draw conclusions.

\section{The particle in a cubic box}

\label{sec:cubic box}

In order to simplify the discussion of the Schr\"{o}dinger equation for the
particle in a cubic box with impenetrable walls we choose the following
units:
\begin{equation}
\;
\begin{array}{ll}
length & \frac{L}{2} \\
energy & \frac{2\hbar ^{2}}{mL^{2}} \\
momentum & \frac{2\hbar }{L}
\end{array}
,  \label{eq:units}
\end{equation}
where $m$ is the mass of the particle and $L$ the length of the box edges.
In this way the Schr\"{o}dinger equation becomes
\begin{eqnarray}
H\psi &=&E\psi ,  \nonumber \\
H &=&p_{x}^{2}+p_{y}^{2}+p_{z}^{2}  \nonumber \\
p_{q} &=&-i\frac{\partial }{\partial q}  \nonumber \\
\psi (\pm 1,y,z) &=&\psi (x,\pm 1,z)=\psi (x,y,\pm 1)=0.
\label{eq:Schrodinger}
\end{eqnarray}
The dimensionless eigenvalues and eigenfunctions are:
\begin{eqnarray}
E_{n_{1}n_{2}n_{3}}
&=&\frac{\pi^2}{4}(n_{1}^{2}+n_{2}^{2}+n_{3}^{2}) \nonumber
\\
\psi _{n_{1}n_{2}n_{3}}(x,y,z) &=&\sin \frac{n_{1}\pi (x+1)}{2}\sin \frac{%
n_{2}\pi (y+1)}{2}\sin \frac{n_{3}\pi (z+1)}{2},
\label{eq:eigenval_eigenfun}
\end{eqnarray}
where the origin of the system of coordinates has been placed at
the center of the box. Note that $\psi
_{n_{1}n_{2}n_{3}}(-x,y,z)=(-1)^{n_{1}+1}\psi
_{n_{1}n_{2}n_{3}}(x,y,z)$ (and similar relationships for the
other two coordinates).

Throughout this paper we resort to the following notation for the
permutation of a set of real numbers
\begin{eqnarray}
\{a,b,b\}_{P} &=&\{\{a,b,b\},\{b,a,b\},\{b,b,a\}\}  \nonumber \\
\{a,b,c\}_{P}
&=&\{\{a,b,c\},\{c,a,b\},\{b,c,a\},\{b,a,c\},\{c,b,a\},\{a,c,b\}\}.
\label{eq:permutations}
\end{eqnarray}
Thus, the set of quantum numbers $\{n,m,m\}_{P}$ leads to a three-fold
degenerate energy level; that is to say, three linearly independent
eigenfunctions with the same energy $E_{nmm}$ (provided that $m\neq n$).
Analogously, three different quantum numbers $\{n_{1},n_{2},n_{3}\}_{P}$
give rise to a six-fold degenerate energy level.

\section{Point-group symmetry}

\label{sec:point-group symmetry}

The suitable point group for describing the symmetry of the
particle is a cubic box is $O_{h}$ shown in Table~\ref{tab:oh}. It
predicts two-fold ($E_{g}$, $E_{u}$) and three-fold ($T_{1g}$,
$T_{2g}$, $T_{1u}$, $T_{2u}$) degenerate energy levels. It is
clear that the geometrical symmetry of the cube is insufficient to
account for the degeneracy already described in the preceding
section. Therefore, as in the case of the particle in a square
box,\cite{LFLA97} there must be a hidden dynamical symmetry. In
order to discuss this point in more detail we should first
classify the eigenfunctions according to their point-group
symmetry.

Such classification is greatly facilitated by the projection operators:\cite
{T64,C90}
\begin{equation}
P^{j}=\frac{l_{j}}{h}\sum_{R}\chi (R)^{j}R  \label{eq:projection}
\end{equation}
where $l_{j}$ is the dimension of the irreducible representation $j$, $h$
the order of the group, $\chi (R)^{j}$ the character of the group operation $%
R$ for the irreducible representation $j$, and the sum is over all the
operations $R$ of the group. The appendix outlines how to obtain the
projection operators.

The first energy levels are
\begin{equation}
\begin{array}{ll}
\{1,1,1\} & A_{1g} \\
\{1,1,2\}_{P} & T_{1u} \\
\{1,2,2\}_{P} & T_{2g} \\
\{1,1,3\}_{P} & E_{g},A_{1g} \\
\{2,2,2\} & A_{2u} \\
\{1,2,3\}_{P} & T_{1u},T_{2u}
\end{array}
,  \label{eq:degeneracy_1}
\end{equation}
and the sets of degenerate eigenfunctions produced by the projection
operators are:
\begin{equation}
\begin{array}{ll}
1A_{1g} & \psi _{111} \\
1T_{1u} & \left\{ \psi _{211},\psi _{121},\psi _{112}\right\}  \\
1T_{2g} & \left\{ \psi _{122},\psi _{212},\psi _{221}\right\}  \\
1E_{g} & \left\{ \frac{1}{3}\left( 2\psi _{311}-\psi _{131}-\psi
_{113}\right) ,\frac{1}{3}\left( \psi _{311}-2\psi _{131}+\psi _{113}\right)
\right\}  \\
2A_{1g} & \frac{1}{3}\left( \psi _{311}+\psi _{131}+\psi _{113}\right)  \\
2T_{1u} & \left\{ \frac{1}{2}\left( \psi _{123}+\psi _{321}\right) ,\frac{1}{%
2}\left( \psi _{312}+\psi _{132}\right) ,\frac{1}{2}\left( \psi _{231}+\psi
_{213}\right) \right\}  \\
1T_{2u} & \left\{ \frac{1}{2}\left( \psi _{123}-\psi _{321}\right) ,\frac{1}{%
2}\left( \psi _{312}-\psi _{132}\right) ,\frac{1}{2}\left( \psi _{231}-\psi
_{213}\right) \right\}
\end{array}
.  \label{eq:eigenf_Oh}
\end{equation}
The two eigenfunctions $E_{g}$ are linearly independent but not orthogonal.
One can easily obtain two orthogonal functions by appropriate linear
combinations. We have just left them as they come from the application of
the projection operator $P^{E_{g}}$. Exactly the same situation takes place
in the case of the degenerate eigenfunctions $2T_{1u}$ and $1T_{2u}$.
Functions of different symmetry are obviously orthogonal. The magnitude of
the energy increases from top to bottom in Eq.~(\ref{eq:eigenf_Oh}). The
three-fold degenerate level given by $\{1,1,3\}_{P}$ with eigenfunctions $%
1E_{g}$ and $2A_{1g}$ and the sixth-fold one $\{1,2,3\}_{P}$ with
eigenfunctions $2T_{1u}$ and $1T_{2u}$ cannot be explained by point-group
symmetry. In general we have
\begin{equation}
\begin{array}{ll}
\{2n-1,2n-1,2n-1\} & A_{1g} \\
\{2n,2n,2n\} & A_{2u} \\
\{2n,2n,2m-1\}_{P} & T_{2g} \\
\{2n-1,2n-1,2m\}_{P} & T_{1u} \\
\{2n-1,2n-1,2m-1\}_{P} & A_{1g},E_{g} \\
\{2n,2n,2m\}_{P} & A_{2u},E_{u} \\
\{2n-1,2m-1,2k-1\}_{P} & A_{1g},A_{2g},E_{g},E_{g} \\
\{2n,2m,2k\}_{P} & A_{1u},A_{2u},E_{u},E_{u} \\
\{2n-1,2m-1,2k\}_{P} & T_{1u},T_{2u} \\
\{2n,2m,2k-1\}_{P} & T_{1g},T_{2g}
\end{array}
.  \label{eq:degeneracy_2}
\end{equation}

In order to understand such degeneracy of the energy levels suppose that
there is an operator $D$ that commutes with $H$ and preserves the boundary
conditions. If $\psi $ is eigenfunction of $H$ with eigenvalue $E$ then $%
D\psi $ is eigenfunction of $H$ with the same eigenvalue: $HD\psi =ED\psi $.
If it happens that $D$ connects functions of different symmetry $S$ and $%
S^{\prime }$ $D\psi ^{S}=\psi ^{S^{\prime }}$ then the degree of degeneracy
is greater than the one predicted by point-group symmetry. The accidental
degeneracy of the energy levels of the particle in a cubic box can be
explained by a couple of operators of symmetry $E_{g}$ as shown by the
products
\begin{eqnarray}
E_{g}\times A_{1g} &=&E_{g}  \nonumber \\
E_{g}\times A_{2g} &=&E_{g}  \nonumber \\
E_{g}\times E_{g} &=&A_{1g}+A_{2g}+E_{g}  \nonumber \\
E_{g}\times T_{1g} &=&T_{1g}+T_{2g}  \nonumber \\
E_{g}\times T_{2g} &=&T_{1g}+T_{2g}  \nonumber \\
E_{g}\times A_{1u} &=&E_{u}  \nonumber \\
E_{g}\times A_{2u} &=&E_{u}  \nonumber \\
E_{g}\times E_{u} &=&A_{1u}+A_{2u}+E_{u}  \nonumber \\
E_{g}\times T_{1u} &=&T_{1u}+T_{2u}  \nonumber \\
E_{g}\times T_{2u} &=&T_{1u}+T_{2u}.  \label{eq:direct_product}
\end{eqnarray}

Application of the projection operator $P^{E_{g}}$ to $x^{2}$ and $y^{2}$
\begin{eqnarray}
P^{E_{g}}x^{2} &=&\frac{1}{3}\left( 2x^{2}-y^{2}-z^{2}\right)  \nonumber \\
P^{E_{g}}y^{2} &=&\frac{1}{3}\left( 2y^{2}-x^{2}-z^{2}\right)
\end{eqnarray}
shows that two suitable operators are
\begin{eqnarray}
D_{E_{g}}(1) &=&2p_{x}^{2}-p_{y}^{2}-p_{z}^{2}  \nonumber \\
D_{E_{g}}(2) &=&2p_{y}^{2}-p_{x}^{2}-p_{z}^{2}.  \label{eq:DEg}
\end{eqnarray}
These are probably the two additional dynamical symmetries
mentioned by Leyvraz et al\cite{LFLA97} and are equivalent to
those derived by Hern\'andez-Castillo and Lemus\cite{HCL13}. We
illustrate the effect of these operators on the particular case
$\{1,3,5\}_{P}$. The functions of symmetry $E_{g}$ are
\begin{eqnarray}
\psi _{E_{g}}^{[1]} &=&2\psi _{135}-\psi _{513}-\psi _{351},  \nonumber \\
\psi _{E_{g}}^{[2]} &=&\psi _{135}-2\psi _{513}+\psi _{351},  \nonumber \\
\psi _{E_{g}}^{[3]} &=&2\psi _{315}-\psi _{531}-\psi _{153},  \nonumber \\
\psi _{E_{g}}^{[4]} &=&\psi _{315}-2\psi _{531}+\psi _{153}.
\end{eqnarray}
The first pair is orthogonal to the second one, but each pair is not
orthogonal as argued above. The application of the operators $D_{E_{g}}$
yields
\begin{eqnarray}
P^{A_{1g}}D_{E_{g}}(1)\psi _{E_{g}}^{[1]} &=&4\pi ^{2}\left( \psi
_{135}+\psi _{513}+\psi _{351}+\psi _{315}+\psi _{531}+\psi _{153}\right)
\nonumber \\
P^{A_{2g}}D_{E_{g}}(1)\psi _{E_{g}}^{[1]} &=&4\pi ^{2}\left( \psi
_{135}+\psi _{513}+\psi _{351}-\psi _{315}-\psi _{531}-\psi _{153}\right)
\nonumber \\
P^{A_{1g}}D_{E_{g}}(1)\psi _{E_{g}}^{[3]} &=&\pi ^{2}\left( \psi _{135}+\psi
_{513}+\psi _{351}+\psi _{315}+\psi _{531}+\psi _{153}\right)  \nonumber \\
P^{A_{2g}}D_{E_{g}}(1)\psi _{E_{g}}^{[3]} &=&-\pi ^{2}\left( \psi
_{135}+\psi _{513}+\psi _{351}-\psi _{315}-\psi _{531}-\psi _{153}\right) .
\end{eqnarray}
We clearly see that the operators $D_{E_{g}}$ already connect the functions
of symmetry $A_{1g}$ and $A_{2g}$ with the two pairs of functions of
symmetry $E_{g}$ and thus account for the six-fold degenerate energy level
with quantum numbers $n_{1}=1$, $n_{2}=3$, $n_{3}=5$. Note that we have
chosen only one member of each pair ($\psi _{E_{g}}^{[1]}$, $\psi
_{E_{g}}^{[3]}$) as an illustrative example. We can prove the other
accidental degeneracies in Eq.~(\ref{eq:degeneracy_2}) exactly in the same
way.

We can build other operators that commute with $H$ and connect functions of
different symmetry. For example,
\begin{equation}
D_{A_{2g}}=\left( p_{x}^{2}-p_{y}^{2}\right) \left(
p_{x}^{2}-p_{z}^{2}\right) \left( p_{y}^{2}-p_{z}^{2}\right) ,
\end{equation}
accounts for the degeneracy of the pairs $\left\{ A_{1g},A_{2g}\right\} $, $%
\left\{ T_{1g},T_{2g}\right\} $, $\left\{ A_{1u},A_{2u}\right\} $, and $%
\left\{ T_{1u},T_{2u}\right\} $. This operator was not mentioned
by Hern\'andez-Castillo and Lemus\cite{HCL13} who certainly
discussed the problem in a more technical way with greater
mathematical rigor that is beyond our more pedagogical aims.

In this paper we do not try to explain the degeneracy that comes
from
\textit{Pythagorean} relations of the form $%
n_{1}^{2}+n_{2}^{2}+n_{3}^{3}=m_{1}^{2}+m_{2}^{2}+m_{3}^{2}$ that
have been already discussed for the square box\cite{L82}. However,
in what follows we show the first cases in increasing energy order
\begin{equation}
\begin{array}{ll}
\{3,3,3\},\,\{1,1,5\}_{P}\, & A_{1g},A_{1g},E_{g} \\
\{1,4,4\}_{P},\,\{2,2,5\}_{P} & T_{2g},T_{2g} \\
\{1,1,6\}_{P},\,\{2,3,5\}_{P} & T_{1u},T_{1u},T_{2u} \\
\{1,2,6\}_{P},\,\{3,4,4\}_{P} & T_{1g},T_{2g},T_{2g} \\
\{1,5,5\}_{P},\,\{1,1,7\}_{P} & A_{1g},E_{g},A_{1g},E_{g} \\
\{2,5,5\}_{P},\,\{3,3,6\}_{P},\,\{1,2,7\}_{P} &
T_{1u},T_{1u},T_{1u},T_{2u}
\end{array}
.
\end{equation}
It seems that the Pythagorean and dynamical symmetries connect the
same kind of irreducible representations.

\section{Perturbation theory}

\label{sec:perturbation theory}

Suppose that we add a potential $V(x,y,z)$ inside the box and obtain
\begin{equation}
H=H_{0}+V,
\end{equation}
where $H_{0}$ is given by Eq.~(\ref{eq:Schrodinger}). In particular we are
interested in the polynomial potential
\begin{equation}
V(x,y,z)=x^{2}y^{2}+x^{2}z^{2}+y^{2}z^{2},  \label{eq:V_Oh}
\end{equation}
that is invariant under the operations of the group $O_{h}$ (note, for
example, that $P^{A_{1g}}V(x,y,z)=V(x,y,z)$). Since $H$ does not commute
with the operators $D$ discussed in the preceding section, then we expect
that the dynamical symmetry is broken and the accidental degeneracy removed.
On the other hand, the point-group symmetry remains unbroken and the
degeneracy of the energy levels is that given by the geometrical symmetry.

Perturbation theory\cite{CDL77,P68,F01} is probably the simplest way of
verifying those conclusions. We write $H=H_{0}+\lambda V$ and expand the
energy in a $\lambda $-power series $E=E^{(0)}+E^{(1)}\lambda +\ldots $.
Straightforward application of this approach yields

\begin{eqnarray}
E_{1A_{1g}} &=&\frac{3}{4}+\lambda \frac{\left( \pi ^{2}-6\right) ^{2}}{3\pi
^{4}}+\ldots  \nonumber \\
E_{1T_{1u}} &=&\frac{3}{2}+\lambda \frac{\left( \pi ^{2}-3\right) \left( \pi
^{2}-6\right) }{3\pi ^{4}}+\ldots  \nonumber \\
E_{1T_{2g}} &=&\frac{9}{4}+\lambda \frac{\left( 2\pi ^{2}-3\right) \left(
2\pi ^{2}-9\right) }{12\pi ^{4}}+\ldots  \nonumber \\
E_{1E_{g}} &=&\frac{11}{4}+\lambda \frac{36\pi ^{4}-304\pi ^{2}+285}{108\pi
^{4}}+\ldots  \nonumber \\
E_{2A_{1g}} &=&\frac{11}{4}+\lambda \frac{18\pi ^{4}-152\pi ^{2}+507}{54\pi
^{4}}+\ldots  \nonumber \\
E_{1A_{2u}} &=&3+\lambda \frac{\left( 2\pi ^{2}-3\right) ^{2}}{12\pi ^{4}}%
+\ldots  \nonumber \\
E_{2T_{2u}} &=&\frac{7}{2}+\lambda \frac{36\pi ^{4}-196\pi ^{2}-75}{108\pi
^{4}}+\ldots  \nonumber \\
E_{2T_{1u}} &=&\frac{7}{2}+\lambda \frac{36\pi ^{4}-196\pi ^{2}+411}{108\pi
^{4}}+\ldots ,  \label{eq:EPT}
\end{eqnarray}
for the lowest eigenvalues. We clearly see that the accidental degeneracy of
the pairs of irreducible representations $\left( E_{g},A_{1g}\right) $ and $%
\left( T_{1u},T_{2u}\right) $ was already broken by the perturbation as
argued above. One can easily carry out the same calculation on the higher
states.

Another interesting perturbation potential is
\begin{equation}
V_{HO}(x,y,z)=x^{2}+y^{2}+z^{2}.  \label{eq:V_HO}
\end{equation}
In this case the resulting Schr\"{o}dinger equation is separable in three
one-dimensional equations of the form
\begin{eqnarray}
\left( p_{q}^{2}+q^{2}\right) \varphi _{n}(q) &=&\epsilon _{n}\varphi
_{n}(q),\,n=1,2,\ldots  \nonumber \\
\varphi _{n}(\pm 1) &=&0,  \label{eq:HO_box}
\end{eqnarray}
and the eigenfunctions and eigenvalues of the whole system are given by
\begin{eqnarray}
\psi _{mnk}(x,y,z) &=&\varphi _{m}(x)\varphi _{n}(y)\varphi _{k}(z)
\nonumber \\
E_{mnk} &=&\epsilon _{m}+\epsilon _{n}+\epsilon _{k}.
\end{eqnarray}
Equation (\ref{eq:HO_box}) cannot be solved exactly but we can obtain
accurate results by means of perturbation theory\cite{F01} or any other
approximate method. For the present purposes it is sufficient to know that
such solution already exists and that $\varphi _{n}(-q)=(-1)^{n+1}\varphi
_{n}(q)$.

It is not difficult to convince oneself that the accidental
degeneracy was not broken by this perturbation, although the
Hamiltonian operator does not commute with the operators
$D_{E_g}(1)$, $D_{E_g}(2)$ and $D_{A_{2g}}$ discussed in
Sec.~\ref{sec:point-group symmetry}. The explanation is that any
function of the operators $H_q=p_{q}^{2}+q^{2}$, $q=x,y,z $,
commutes with $H$ and we can therefore construct similar operators
$D_S$ by simply substituting $H_q$ for $p_{q}^{2}$ in the
expressions derived in Sec.~\ref{sec:point-group symmetry}. For
example,
\begin{eqnarray}
D_{E_{g}}(1) &=&2H_{x}-H_{y}-H_{z}  \nonumber \\
D_{E_{g}}(2) &=&2H_{y}-H_{x}-H_{z},
\end{eqnarray}
are two operators of symmetry $E_g$ that commute with $H$.

The discussion of the three-dimensional oscillator in a cubic box suggests a
straightforward generalization. If we have a Hamiltonian operator of the
form
\begin{equation}
H=\sum_{j=1}^{M}H_{j},
\end{equation}
where $[H_{j},H_{k}]=0$, then we can construct $M-1$ operators of the form
\begin{equation}
D_{k}=\sum_{j=1}^{M-1}d_{kj}H_{j},\,k=1,2,\ldots ,M-1,
\end{equation}
that are linearly independent and commute with $H$. It may be possible that
a judicious choice of the coefficients $d_{kj}$ leads to operators that
connect functions of different symmetry. The number of operators that we can
obtain in this way is enormous because any function of the operators $H_{j}$
will commute with $H$ and we expect some kind of dynamical symmetry emerging
from it. We have already seen some examples in the preceding section.

\section{Conclusions}

\label{sec:conclusions}

We have shown that the point-group symmetry is insufficient to
account for the degree of degeneracy of the energy levels of the
particle in a cubic box. The additional degeneracy is due to a
dynamical symmetry given by operators that commute with the
Hamiltonian one. We have shown how to derive such operators by
means of the projection operators of the group $O_{h}$. The next
step would be to build a larger group that embodies both the
point-group operations as well as the dynamical operators. We do
not do it here because we want to keep this paper as simple as
possible. We refer the reader to the more technical paper by
Hern\'andez-Castillo and Lemus\cite{HCL13}. It is clear that the
particle in two and three-dimensional boxes are suitable exactly
solvable problems for teaching the occurrence of both geometrical
and dynamical types of symmetry.

The cubic box is somewhat more complicated than the square one. In the case
of the square box one can obtain the linear combinations of eigenfunctions
that are bases for the irreducible representations by inspection. In the
case of the cubic box it is preferable to resort to a more systematic
approach based on projection operators. This procedure is straightforward
but rather cumbersome for hand calculation and it is therefore convenient to
resort to computer algebra to speed it. For this reason, the particle in a
cubic box is a suitable example to encourage the students to get some skills
in group theory as well as in computer algebra. In the appendix we outline
how to construct the matrix representation of the group operations as well
as the projection operators.

\begin{table}[]
{\tiny \caption{Character table for $O_h$ point group}
\label{tab:oh}
\begin{tabular}{l|rrrrrrrrrr|l|l}

$O_h$ & $E$ & $8C_3$ & $6C_2$ & $6C_4$ & $3C_2(=C_4^2)$ & $i$ & $6S_4$ & $8S_6$ & $3\sigma_h$ & $6\sigma_d$ \\
\hline
A1g &  1 &   1  &  1  &  1  &  1  &  1  &  1  &  1 &   1  &  1        & &                     $x^2+y^2+z^2$  \\
A2g &  1 &   1  &  -1 &  -1 &  1  &  1  &  -1 &  1 &   1  &  -1      & &                                    \\
Eg  &  2 &   -1 &  0  &  0  &  2  &  2  &  0  &  -1&   2  &  0       & &          $(2z^2-x^2-y^2, x^2-y^2)$  \\
T1g &  3 &   0  &  -1 &  1  &  -1 &  3  &  1  &  0 &   -1 &  -1      & $(R_x, R_y, R_z)$&           \\
T2g &  3 &   0  &  1  &  -1 &  -1 &  3  &  -1 &  0 &   -1 &  1       & $(xz, yz, xy)$   &              \\
A1u &  1 &   1  &  1  &  1  &  1  &  -1 &  -1 &  -1&   -1 &  -1     &&         \\
A2u &  1 &   1  &  -1 &  -1 &  1  &  -1 &  1  &  -1&   -1 &  1      &&         \\
Eu  &  2 &   -1 &  0  &  0  &  2  &  -2 &  0  &  1 &   -2 &  0      &&         \\
T1u &  3 &   0  &  -1 &  1  &  -1 &  -3 &  -1 &  0 &   1  &  1      & $(x, y, z)$ &               \\
T2u &  3 &   0  &  1  &  -1 &  -1 &  -3 &  1  &  0 &   1  &  -1      &&         \\

\end{tabular}
}
\end{table}

\appendix

\section{Symmetry operations, matrix representation and projection
operators}

The analysis of the particle in a cubic box presented in Sec.~\ref
{sec:point-group symmetry} is greatly facilitated by the application of
projection operators (\ref{eq:projection}). For this purpose we need to know
the effect of the symmetry operations on the cartesian coordinates $\mathbf{x%
}$ that we express in matrix form as $\mathbf{x}^{\prime
}=\mathbf{Mx}$, where $\mathbf{M}$ is a $3\times 3$ unitary matrix
and $\mathbf{x}$ and $\mathbf{x}^{\prime }$ are $3\times 1$ column
matrices. Once we have the matrix representation $\mathbf{M}$ for
a given symmetry operation $R$ then we easily derive the effect of
the latter on a function $f(\mathbf{x})$ as\cite{T64}
\begin{equation}
Rf(x)=f\left( \mathbf{M}^{-1}\mathbf{x}\right) .  \label{eq:Rf}
\end{equation}
Taking into account equations (\ref{eq:projection}) and (\ref{eq:Rf}) the
application of the projection operator $P^{j}$ on $f(\mathbf{x})$ is
straightforward.

The symmetry elements of the group $O_{h}$ are summarized in most
books on group theory\cite{C90}. The matrix representation of the
48 symmetry operations is given explicitly in a recent paper by
Delibas et al\cite {DATA13}. For the present study of the particle
in a cubic box we built the matrices without locating all the
symmetry elements of the cube explicitly. We resorted to a rather
more algebraic procedure that we describe in what follows because
it may be useful for those who prefer a less geometrical approach.

To begin with we obtain the 48 matrices for the following coordinate
transformations (note that each line embodies 6 transformations)
\begin{eqnarray}
\{x,y,z\} &\rightarrow &\{x,y,z\}_{P}  \nonumber \\
\{x,y,z\} &\rightarrow &\{-x,y,z\}_{P}  \nonumber \\
\{x,y,z\} &\rightarrow &\{x,-y,z\}_{P}  \nonumber \\
\{x,y,z\} &\rightarrow &\{x,y,-z\}_{P}  \nonumber \\
\{x,y,z\} &\rightarrow &\{-x,-y,z\}_{P}  \nonumber \\
\{x,y,z\} &\rightarrow &\{-x,y,-z\}_{P}  \nonumber \\
\{x,y,z\} &\rightarrow &\{x,-y,-z\}_{P}  \nonumber \\
\{x,y,z\} &\rightarrow &\{-x,-y,-z\}_{P}.
\end{eqnarray}
The next step is to identify the symmetry operation associated to each
matrix. First, note that the traces of these matrices are the characters for
the $T_{1u}$ irreducible representation. Second, take into account that the
determinant of a rotation $C_{n}$ is unity and the determinants of a
reflection $\sigma $ and the improper rotation $S_{n}$ are minus one. Third,
remember that the order $n$ of a symmetry operation $a$ is the smallest
positive integer such that $a^{n}=E$ (the identity operation). For example, $%
\det \mathbf{M}(S_{6})=-1$ and $\mathbf{M}(S_{6})^{6}=\mathbf{I}$
(the identity matrix). We can thus group all the matrices derived
above in the corresponding group classes, except $6\ C_{2}$ and
$3\ C_{2}(=C_{4}^{2})$\cite {C90} that share the same trace,
determinant, and order. The identification of the matrices for the
three rotations by an angle $\pi $ about the coordinate axes
($C_{2}=C_{4}^{2}$) is straightforward and the remaining six
matrices represent the rotations $C_{2}$ about axes that bisect
opposite edges of the cube. In this way we obtain

\begin{equation}
\mathbf{E}=\left(
\begin{array}{rrr}
1 & 0 & 0 \\
0 & 1 & 0 \\
0 & 0 & 1
\end{array}
\right) ,\;\mathbf{i}=\left(
\begin{array}{rrr}
-1 & 0 & 0 \\
0 & -1 & 0 \\
0 & 0 & -1
\end{array}
\right)
\end{equation}

$8C_{3}$
\begin{eqnarray}
&&\left(
\begin{array}{rrr}
0 & 1 & 0 \\
0 & 0 & 1 \\
1 & 0 & 0
\end{array}
\right) ,\left(
\begin{array}{rrr}
0 & 0 & 1 \\
1 & 0 & 0 \\
0 & 1 & 0
\end{array}
\right) ,\left(
\begin{array}{rrr}
0 & -1 & 0 \\
0 & 0 & -1 \\
1 & 0 & 0
\end{array}
\right) ,\left(
\begin{array}{rrr}
0 & 0 & -1 \\
-1 & 0 & 0 \\
0 & 1 & 0
\end{array}
\right)  \nonumber \\
&&\left(
\begin{array}{rrr}
0 & 1 & 0 \\
0 & 0 & -1 \\
-1 & 0 & 0
\end{array}
\right) ,\left(
\begin{array}{rrr}
0 & 0 & 1 \\
-1 & 0 & 0 \\
0 & -1 & 0
\end{array}
\right) ,\left(
\begin{array}{rrr}
0 & -1 & 0 \\
0 & 0 & 1 \\
-1 & 0 & 0
\end{array}
\right) ,\left(
\begin{array}{rrr}
0 & 0 & -1 \\
1 & 0 & 0 \\
0 & -1 & 0
\end{array}
\right)  \nonumber \\
\end{eqnarray}

$6C_2$

\begin{eqnarray}
\left(
\begin{array}{rrr}
-1 & 0 & 0 \\
0 & 0 & 1 \\
0 & 1 & 0
\end{array}
\right),\left(
\begin{array}{rrr}
0 & 0 & 1 \\
0 & -1 & 0 \\
1 & 0 & 0
\end{array}
\right),\left(
\begin{array}{rrr}
0 & 1 & 0 \\
1 & 0 & 0 \\
0 & 0 & -1
\end{array}
\right),  \nonumber \\
\left(
\begin{array}{rrr}
-1 & 0 & 0 \\
0 & 0 & -1 \\
0 & -1 & 0
\end{array}
\right),\left(
\begin{array}{rrr}
0 & 0 & -1 \\
0 & -1 & 0 \\
-1 & 0 & 0
\end{array}
\right),\left(
\begin{array}{rrr}
0 & -1 & 0 \\
-1 & 0 & 0 \\
0 & 0 & -1
\end{array}
\right)
\end{eqnarray}

$6C_4$

\begin{eqnarray}
\left(
\begin{array}{rrr}
0 & -1 & 0 \\
1 & 0 & 0 \\
0 & 0 & 1
\end{array}
\right),\left(
\begin{array}{rrr}
0 & 0 & -1 \\
0 & 1 & 0 \\
1 & 0 & 0
\end{array}
\right),\left(
\begin{array}{rrr}
0 & 1 & 0 \\
-1 & 0 & 0 \\
0 & 0 & 1
\end{array}
\right),  \nonumber \\
\left(
\begin{array}{rrr}
1 & 0 & 0 \\
0 & 0 & -1 \\
0 & 1 & 0
\end{array}
\right),\left(
\begin{array}{rrr}
0 & 0 & 1 \\
0 & 1 & 0 \\
-1 & 0 & 0
\end{array}
\right),\left(
\begin{array}{rrr}
1 & 0 & 0 \\
0 & 0 & 1 \\
0 & -1 & 0
\end{array}
\right)
\end{eqnarray}

$3C_2(=C_4^2)$

\begin{equation}
\left(
\begin{array}{rrr}
-1 & 0 & 0 \\
0 & -1 & 0 \\
0 & 0 & 1
\end{array}
\right) ,\left(
\begin{array}{rrr}
1 & 0 & 0 \\
0 & -1 & 0 \\
0 & 0 & -1
\end{array}
\right) ,\left(
\begin{array}{rrr}
-1 & 0 & 0 \\
0 & 1 & 0 \\
0 & 0 & -1
\end{array}
\right)
\end{equation}

$6S_4$

\begin{eqnarray}
\left(
\begin{array}{rrr}
-1 & 0 & 0 \\
0 & 0 & -1 \\
0 & 1 & 0
\end{array}
\right),\left(
\begin{array}{rrr}
0 & 0 & -1 \\
0 & -1 & 0 \\
1 & 0 & 0
\end{array}
\right),\left(
\begin{array}{rrr}
0 & 0 & 1 \\
0 & -1 & 0 \\
-1 & 0 & 0
\end{array}
\right),  \nonumber \\
\left(
\begin{array}{rrr}
0 & 1 & 0 \\
-1 & 0 & 0 \\
0 & 0 & -1
\end{array}
\right),\left(
\begin{array}{rrr}
-1 & 0 & 0 \\
0 & 0 & 1 \\
0 & -1 & 0
\end{array}
\right),\left(
\begin{array}{rrr}
0 & -1 & 0 \\
1 & 0 & 0 \\
0 & 0 & -1
\end{array}
\right)
\end{eqnarray}

$8S_6$

\begin{eqnarray}
&&\left(
\begin{array}{rrr}
0 & -1 & 0 \\
0 & 0 & 1 \\
1 & 0 & 0
\end{array}
\right) ,\left(
\begin{array}{rrr}
0 & 0 & -1 \\
1 & 0 & 0 \\
0 & 1 & 0
\end{array}
\right) ,\left(
\begin{array}{rrr}
0 & 1 & 0 \\
0 & 0 & -1 \\
1 & 0 & 0
\end{array}
\right) ,\left(
\begin{array}{rrr}
0 & 0 & 1 \\
-1 & 0 & 0 \\
0 & 1 & 0
\end{array}
\right)  \nonumber \\
&&\left(
\begin{array}{rrr}
0 & 1 & 0 \\
0 & 0 & 1 \\
-1 & 0 & 0
\end{array}
\right) ,\left(
\begin{array}{rrr}
0 & 0 & 1 \\
1 & 0 & 0 \\
0 & -1 & 0
\end{array}
\right) ,\left(
\begin{array}{rrr}
0 & -1 & 0 \\
0 & 0 & -1 \\
-1 & 0 & 0
\end{array}
\right) ,\left(
\begin{array}{rrr}
0 & 0 & -1 \\
-1 & 0 & 0 \\
0 & -1 & 0
\end{array}
\right)  \nonumber \\
\end{eqnarray}

$3\sigma_h$

\begin{equation}
\left(
\begin{array}{rrr}
-1 & 0 & 0 \\
0 & 1 & 0 \\
0 & 0 & 1
\end{array}
\right) ,\left(
\begin{array}{rrr}
1 & 0 & 0 \\
0 & -1 & 0 \\
0 & 0 & 1
\end{array}
\right) ,\left(
\begin{array}{rrr}
1 & 0 & 0 \\
0 & 1 & 0 \\
0 & 0 & -1
\end{array}
\right)
\end{equation}

$6\sigma_d$

\begin{eqnarray}
\left(
\begin{array}{rrr}
1 & 0 & 0 \\
0 & 0 & 1 \\
0 & 1 & 0
\end{array}
\right),\left(
\begin{array}{rrr}
0 & 0 & 1 \\
0 & 1 & 0 \\
1 & 0 & 0
\end{array}
\right),\left(
\begin{array}{rrr}
0 & 1 & 0 \\
1 & 0 & 0 \\
0 & 0 & 1
\end{array}
\right),  \nonumber \\
\left(
\begin{array}{rrr}
0 & -1 & 0 \\
-1 & 0 & 0 \\
0 & 0 & 1
\end{array}
\right),\left(
\begin{array}{rrr}
1 & 0 & 0 \\
0 & 0 & -1 \\
0 & -1 & 0
\end{array}
\right),\left(
\begin{array}{rrr}
0 & 0 & -1 \\
0 & 1 & 0 \\
-1 & 0 & 0
\end{array}
\right)
\end{eqnarray}

\end{document}